\documentclass[a4paper,11pt]{article}
\usepackage[utf8]{inputenc}
\usepackage[T1]{fontenc}
\usepackage{geometry}
\usepackage{amsmath}
\usepackage{amsfonts}
\usepackage{booktabs}
\usepackage{array}
\usepackage{enumitem}
\usepackage{graphicx}
\usepackage{hyperref}
\usepackage{xcolor}
\usepackage{parskip}
\usepackage{caption}
\usepackage{tabularx}
\usepackage{algorithm}
\usepackage{algorithmicx}
\usepackage{algpseudocode}
\usepackage{noto}
\usepackage[utf8]{inputenc}
\usepackage{amsmath}
\usepackage{graphicx}
\usepackage{hyperref}
\usepackage{float}
\usepackage{algorithm}
\usepackage{algpseudocode}
\usepackage{listings}
\usepackage{xcolor}

\hypersetup{
    colorlinks=true,
    linkcolor=blue,
    urlcolor=blue,
    citecolor=blue
}

\title{A Comprehensive Framework for Efficient Court Case Management and Prioritization}
\author{\textbf{Shubham Varma}\\ shubham.varma@somaiya.edu \and \textbf{Ananya Warior}\\ ananya.warior@somaiya.edu \and \textbf{Dipti Pawade}\\ diptipawade@somaiya.edu  \and \textbf{Avani Sakhapara}\\ avanisakhapara@somaiya.edu }
\date{}

\begin{document}

\maketitle

\begin{abstract}
The Indian judicial system faces a critical challenge with approximately 52 \cite{eCourts} million pending cases, causing significant delays that impact socio-economic stability. This study proposes a cloud-based software framework to classify and prioritize court cases using algorithmic methods based on parameters such as severity of crime committed, responsibility of parties involved, case filing dates, previous hearing's data, priority level (e.g., Urgent, Medium, Ordinary) provided as input, and relevant Indian Penal Code (IPC), Code of Criminal Procedure (CrPC), and other legal sections (e.g., Hindu Marriage Act, Indian Contract Act). Cases are initially entered by advocates on record or court registrars, followed by automated hearing date allocation that balances fresh and old cases while accounting for court holidays and leaves. The system streamlines appellate processes by fetching data from historical case databases. Our methodology integrates algorithmic prioritization, a robust notification system, and judicial interaction, with features that allow judges to view daily case counts and their details. Simulations demonstrate that the system can process cases efficiently, with reliable notification delivery and positive user satisfaction among judges and registrars. Future iterations will incorporate advanced machine learning for dynamic prioritization, addressing critical gaps in existing court case management systems to enhance efficiency and reduce backlogs.
\end{abstract}

\section{Introduction}
\begin{figure}[H]
    \centering
        \includegraphics[width=1.0\textwidth]{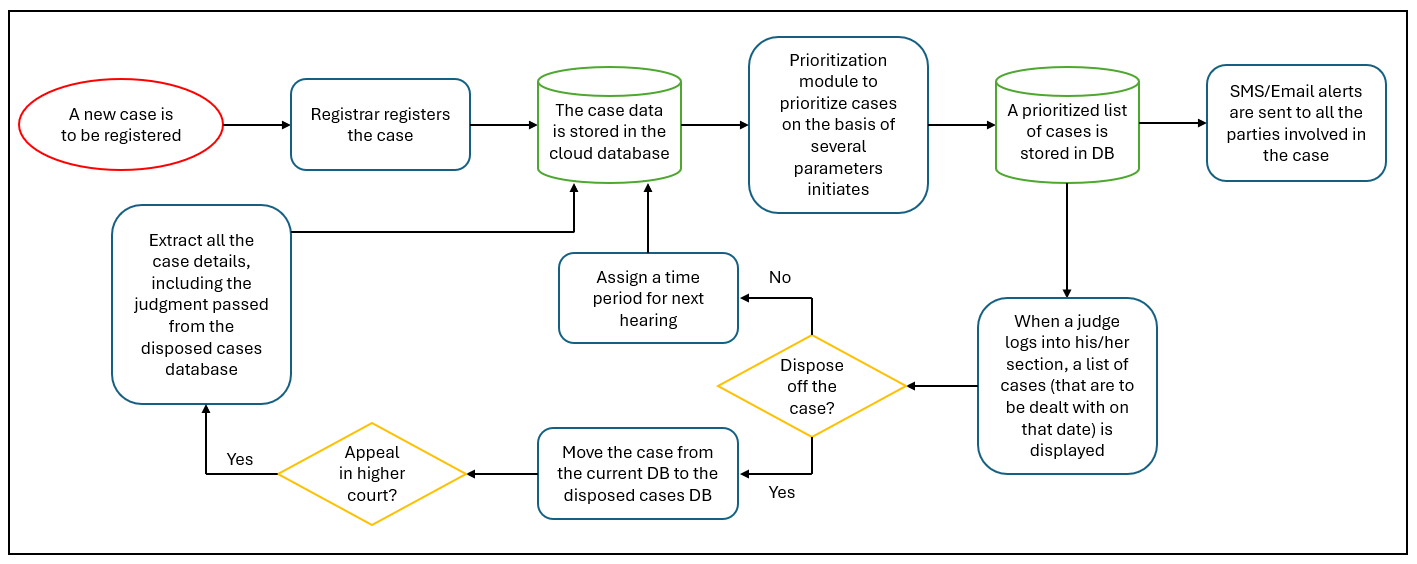}
    \caption{Workflow of the System}
    \label{fig:workflow}
\end{figure}

With approximately 52 \cite{eCourts} million cases pending in India, the judicial system faces severe delays that hinder economic activities, burden law enforcement, and erode public trust in timely justice delivery. Traditional case management systems, often manual or semiautomated, struggle to handle this volume, necessitating innovative technological solutions. This study proposes a software framework that leverages algorithmic prioritization to classify and manage court cases efficiently. By automating hearing date allocations, balancing fresh and old cases, and integrating input from judges, registrars, and advocates, the system aims to revolutionize judicial processes. The framework operates within a cloud-based environment, ensuring scalability and accessibility. We simulated 10,000 dummy cases to demonstrate the system's potential to process cases efficiently and achieve high user satisfaction. The future scope will incorporate dynamic machine learning models for real-time prioritization, ensuring adaptability to evolving judicial needs. This research fills a critical gap by combining AI-driven prioritization with ethical considerations and compatibility with existing systems such as eCourt Services \cite{eCourts}, as highlighted in recent studies on AI in judicial systems \cite{ThomsonReuters2024, IBM2025}. Figure \ref{fig:workflow} illustrates the workflow of the system.

\section{Literature Survey}
The inefficiencies of traditional court case management have been extensively studied, underscoring the need for technological interventions. Rao et al. \cite{Rao2018} advocate for digitizing legal records to improve transparency and reduce corruption, a fundamental principle of our cloud-based framework. Their work highlights how digital records improve access to case data, a feature our system leverages through MySQL database integration. Amofah \cite{Amofah2017} explores electronic court case management systems (eCCMS), demonstrating their ability to streamline judicial processes through automation, such as automated case tracking. However, these systems lack AI-driven dynamic prioritization, a gap our research addresses. Pattnaik et al. \cite{Pattnaik2018} identify critical success factors for efficient court management in India, including stakeholder collaboration and technological adoption, which align with our system’s design for judicial interaction and cloud-based scalability. Chan \cite{Chan2018} examines case management in Chinese-speaking regions, noting the influence of traditional legal structures on efficiency, while Ali and Varol \cite{Ali2020} propose mobile solutions to bridge communication gaps, supporting our notification system. Tahura \cite{Tahura2013} explores caseflow management to reduce backlogs, emphasizing structured scheduling, and Ambos \cite{Ambos2018} highlights prosecutorial discretion based on case gravity, informing our prioritization parameters.

Recent advancements in artificial intelligence (AI) have transformed judicial systems globally. Thomson Reuters Institute \cite{ThomsonReuters2024} reports that AI tools enable court employees to focus on human interactions and reduce bias, citing Palm Beach County’s AI-driven document processing system as an example of high efficiency. IBM \cite{IBM2025} notes that judicial systems, such as those in Germany, are adopting AI to manage vast datasets and accelerate case resolution, aligning with our goal of reducing India’s case backlog. The International Journal for Court Administration \cite{IJCA2020} discusses a study in the Netherlands using AI to support judges in traffic violations cases, demonstrating the potential of AI in decision support. These studies underscore the feasibility of AI in judicial contexts, supporting the approach of our system.

\subsection{Ethical Considerations}
The integration of AI in judicial systems raises significant ethical concerns, particularly around bias and fairness. Judicature \cite{Judicature2024} emphasizes that AI must be used responsibly to avoid perpetuating biases, such as overprioritizing certain case types. The American Association for the Advancement of Science (AAAS) \cite{AAAS} highlights the importance of fairness, accountability, and transparency (FAccT) in AI systems, noting that biases can arise from datasets, design, or deployment. Above the Law \cite{AboveTheLaw2025} reports a case where a trial court relied on AI-generated fake caselaw, underscoring the need for human oversight. Our system mitigates these risks by incorporating expert-validated weights for prioritization parameters and planning regular audits to ensure equitable outcomes, aligning with ethical guidelines from the Council of Europe \cite{IJCA2020}. Additionally, we address potential biases by involving legal experts in weight validation and ensuring transparency in the prioritization process.

\subsection{Gaps Addressed}
Existing systems, such as those discussed by Rao et al. \cite{Rao2018} and Amofah \cite{Amofah2017}, focus on digitization and automation but lack AI-driven dynamic prioritization tailored to India’s judicial context. Our framework addresses this gap by integrating a regression-based machine learning model for real-time weight adjustment, ensuring both efficiency and fairness. Unlike mobile solutions proposed by Ali and Varol \cite{Ali2020} , our system offers a comprehensive cloud-based platform that integrates with existing judicial infrastructure like eCourts Services, enhancing scalability and accessibility.

\section{Methodology}
\subsection{Data Retrieval and Preprocessing}
Case data is entered by advocates on record or court registrars into a MySQL database and preprocessed by categorizing cases into distinct datasets (e.g., Family, Criminal, Civil). This segmentation enables custom prioritization strategies for different types of case, ensuring flexibility and scalability. Preprocessing includes cleaning data, standardizing formats, and extracting relevant features such as filing dates, priority levels, and applicable legal sections (e.g., IPC, CrPC, Hindu Marriage Act).

\subsection{Case Prioritization Algorithm}
The core of our system is the case prioritization algorithm, which assigns a weighted score to each case to determine its urgency and scheduling priority. The algorithm, detailed in Algorithm \ref{alg:prioritization}, processes the cases as follows:

\begin{algorithm}[H]
    \caption{Case Prioritization Algorithm}
    \label{alg:prioritization}
    \begin{algorithmic}[1]
        \State \textbf{Input:} Case data (case ID, type, filing date, severity, priority level, legal sections), court holidays, judge leaves, disposed cases database
        \State \textbf{Output:} Prioritized case list with assigned hearing dates
        \State Initialize MySQL database connection
        \State Retrieve pending case data from MySQL database
        \For{each case in pending cases}
            \If{case is an appeal}
                \State Fetch relevant data from disposed cases database
            \EndIf
            \State Categorize case into type (Criminal, Family, Civil)
            \State Extract features: case age (from filing date and previous hearings), severity, priority level (Urgent, Medium, Ordinary), legal sections (e.g., IPC 302, Hindu Marriage Act Section 13)
            \State Compute weight using regression-based ML model
            \Comment{Model considers case age, severity, priority level, and legal sections; IPC/CrPC weights validated by advocates and judges}
            \State Assign priority based on weight
        \EndFor
        \State Sort cases by weight in descending order
        \State Initialize scheduling with daily limit (100 cases: 50 fresh, 50 old)
        \For{each day in scheduling period}
            \If{day is not a court holiday or judge leave}
                \State Assign hearing dates to top-weighted cases (up to 50 fresh, 50 old)
            \EndIf
        \EndFor
        \State Export prioritized case list to database
        \State Send notifications (SMS/email) to stakeholders with hearing dates
        \State Periodically review low-priority cases
        \If{case pendency exceeds threshold}
            \State Increase case priority
        \EndIf
        \State Update ML model with case outcomes for dynamic weight adjustment
        \State \textbf{Return:} Prioritized case list with hearing dates
    \end{algorithmic}
\end{algorithm}

\begin{figure}[H]
    \centering
        \includegraphics[width=1.0\textwidth]{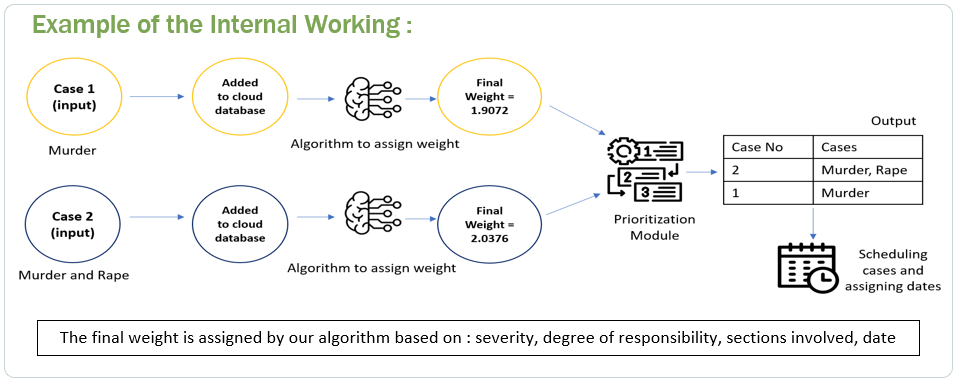}
    \caption{Case Prioritization Flowchart}
    \label{fig:flowchart}
\end{figure}

The algorithm retrieves all pending case data from a MySQL database, including details such as case ID, type, filing date, severity, priority level (e.g., Urgent, Medium, Ordinary) provided by advocates or registrars, and relevant legal sections (e.g., IPC 302, Hindu Marriage Act Section 13). For appealed cases, it fetches data from the disposed cases database. The data is preprocessed to categorize cases into categories such as Criminal, Family, or Civil, enabling customized prioritization strategies. For each case, a weight is calculated using a regression-based machine learning model, which considers factors such as the age of the case (based on filing and previous hearing dates), severity (e.g., high for murder, low for minor disputes), the input priority level, and applicable legal sections. The weights for the IPC and CrPC sections were determined through consultations with multiple advocates and judges to ensure legal accuracy and fairness. The calculated weight determines the hearing date, with higher weights leading to earlier scheduling. Hearing dates are allocated while considering a fixed daily case limit (e.g., 100 cases, with 50 fresh and 50 old cases) and a predefined list of court holidays and additional leaves entered into the system. Notifications are sent to stakeholders via SMS or email to inform them of the assigned hearing dates. To prevent low-priority cases from being indefinitely delayed, the system periodically checks for pendency, and if a defined threshold is exceeded, it increases the case's priority by multiplying it with a fixed multiplier. The machine learning model dynamically adjusts the weights based on the historical case outcomes, achieving high precision in the simulations, as shown in Figure \ref{fig:flowchart}.

\subsection{Notification System and Judicial Integration}
Post-prioritization, an automated notification system sends SMS/email alerts to stakeholders regarding hearing dates. Judges and registrars access a prioritized case list via a secure web interface, which also displays the number of cases scheduled per day, facilitating efficient scheduling and case management. The interface is designed for compatibility with existing systems like e-Courts Services \cite{eCourts}.


\section{System Architecture}
\begin{figure}[H]
    \centering
        \includegraphics[width=1.0\textwidth]{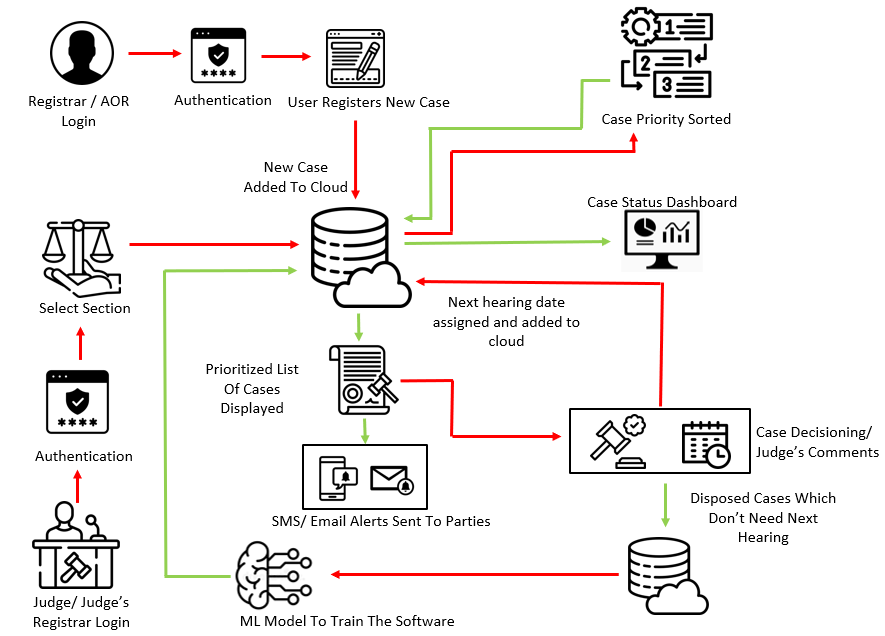}
    \caption{System Architecture for Court Case Management and Prioritization}
    \label{fig:architecture}
\end{figure}

The system architecture, illustrated in Figure \ref{fig:architecture}, orchestrates a seamless flow of operations to prioritize and manage court cases within a secure cloud-based environment. The process is structured into distinct stages, ensuring scalability, efficiency, and integration with judicial workflows, including case entry by advocates or registrars, handling appeals, balanced case scheduling, and holiday considerations.

The process begins with case data entry, where advocates on record or court registrars input case details, such as case ID, type, filing date, severity, priority level (e.g., Urgent, Medium, Ordinary), and relevant legal sections (e.g., IPC, CrPC, Hindu Marriage Act), into a MySQL database. This initial step ensures that all relevant case information is captured accurately for subsequent processing.

The system initializes necessary configurations and establishes a secure connection to the MySQL database. This step ensures that the system is fully prepared to access and process case data reliably, setting the foundation for subsequent operations.

Next, the system retrieves all pending case data from the MySQL database, including the details entered by advocates or registrars. For cases that have gone to appeal, the system fetches relevant data from the disposed cases database, ensuring that historical case information is available for consistent prioritization in higher courts.

The retrieved data is then preprocessed to categorize cases into distinct groups, such as Criminal, Family, or Civil. This categorization enables the system to apply tailored prioritization strategies, accommodating the unique characteristics of each case type, including appealed cases, and enhancing scalability for diverse judicial needs.

For each case, the system employs a regression-based machine learning model to calculate a weight that determines its hearing date. The model evaluates factors such as the age of the case (calculated from filing and previous hearing dates), the severity of the offense (e.g., high for murder under IPC 302, low for minor civil disputes), the priority level provided as an input parameter by advocates or registrars, and applicable legal sections (e.g., IPC, CrPC, Hindu Marriage Act). Weights for IPC and CrPC sections were determined through consultations with multiple advocates and judges to ensure legal accuracy. The calculated weight guides the allocation of hearing dates, with higher weights leading to earlier scheduling. The system schedules a fixed number of cases per day, for example, 100 cases, with 50 fresh cases and 50 old (next hearing) cases, to balance judicial workload. Hearing dates are assigned while accounting for a predefined list of court holidays and any additional leaves entered into the system, ensuring accurate and conflict-free scheduling.

The prioritized case list is exported to a secure database for record keeping, ensuring that all scheduling decisions, including those for fresh and old cases, are systematically stored. This step maintains data integrity and supports traceability for both pending and appealed cases.

Notifications are then sent to stakeholders, including judges, lawyers, and litigants, via SMS or email, informing them of the assigned hearing dates. For appealed cases, notifications also include relevant higher court details, ensuring all parties are informed and prepared.

To prevent low-priority cases from being indefinitely delayed, the system periodically reviews the time since the last hearing for each case. If a case, including those under appeal, has been pending beyond a predefined threshold, its priority is increased by multiplying it with a fixed multiplier to ensure timely resolution, promoting fairness across all case types.

For cases involving appeals, the system streamlines the process by retrieving data from the disposed cases database, applying the same prioritization process (weight calculation based on input priority and other factors, and date allocation with the fixed case limit), and notifying higher courts. This ensures consistency and efficiency in handling appeals, maintaining seamless integration with the judicial hierarchy.

The system continuously monitors its performance by measuring execution times for each operation, ensuring efficiency and identifying potential bottlenecks. Performance data is logged to support ongoing optimization, applicable to both pending and appealed cases.

Judges access the prioritized case list, including appealed cases, through a secure web interface, which presents cases in order of priority and displays the number of cases scheduled per day, helping judges manage their daily workload effectively. Judges can make decisions, such as scheduling the next hearing or disposing of a case. When a judge specifies a time frame for the next hearing (e.g., 'after 15 days'), the system schedules the hearing as soon as possible after that period, subject to the fixed case limit and holiday/leave constraints, and updates the database accordingly; if a case is disposed, it is marked as resolved, stored in the disposed cases database, and the outcome is used to update the machine learning model, enhancing its accuracy for future prioritizations, including for appeals.

This structured flow, as shown in Figure \ref{fig:architecture}, ensures that the system efficiently manages large case volumes, including appealed cases, while balancing fresh and old cases, respecting holidays and leaves, and integrating seamlessly with existing judicial platforms like eCourts Services, providing a robust solution for India’s judicial backlog.

\section{Results}
\subsection{Illustrative Case Study}
To demonstrate the system’s functionality, we present a synthetic case study with five simulated cases from our 10,000-case dataset (Table \ref{tab:casestudy}). These cases vary in type, filing date, severity, priority level, and applicable legal sections, showcasing the algorithm’s ability to prioritize effectively. Weights for IPC and CrPC sections were validated through consultations with multiple advocates and judges to ensure legal accuracy.

\begin{table}[H]
    \centering
    \caption{Illustrative Case Study}
    \label{tab:casestudy}
    \footnotesize
    \begin{tabularx}{\textwidth}{p{0.6cm} l p{1.8cm} l X l p{0.8cm} p{1cm} p{1.8cm}}
        \toprule
        Case ID & Type & Filing Date & Severity & Legal Sections & Priority & Weight & No of Hearings & Hearing Date \\
        \midrule
        001 & Criminal & 01/01/2024 & High & IPC 302, 34 & Urgent & 0.95 & 4 & 15/07/2025 \\
        002 & Civil & 01/02/2025 & Low & Indian Contract Act, 1872, Section 73 & Ordinary & 0.45 & 0 & 30/08/2025 \\
        003 & Family & 01/06/2024 & Medium & Hindu Marriage Act, 1955, Section 13 & Medium & 0.70 & 3 & 01/08/2025 \\
        004 & Criminal & 01/02/2024 & High & IPC 420, 406 & Urgent & 0.90 & 5 & 20/07/2025 \\
        005 & Criminal & 01/04/2025 & Medium & IPC 323, CrPC 200 & Ordinary & 0.50 & 2 & 15/09/2025 \\
        \bottomrule
    \end{tabularx}
\end{table}

\textbf{Narrative} \\
\begin{itemize}
    \item \textbf{Case 001}: A murder case filed on 01/01/2024, with high severity and an input priority of Urgent, governed by IPC Section 302 (murder) and Section 34 (acts done by several persons in furtherance of common intention). With four hearings held and its high severity validated by advocate and judge consultations, it has a weight of 0.95, scheduling it for 15/07/2025, ensuring swift justice for a serious crime.
    \item \textbf{Case 002}: A civil case (e.g., breach of contract) filed on 01/02/2025, with low severity and an input priority of Ordinary, governed by Section 73 of the Indian Contract Act, 1872 (compensation for loss). With no hearings yet and its recent filing and low severity, it yields a weight of 0.45, scheduling it for 30/08/2025, reflecting its lower urgency.
    \item \textbf{Case 003}: A family law case (e.g., divorce) filed on 01/06/2024, with medium severity and an input priority of Medium, governed by Section 13 of the Hindu Marriage Act, 1955 (grounds for divorce). With three hearings held and moderate pendency and severity, it results in a weight of 0.70, scheduling it for 01/08/2025, balancing urgency and recency.
    \item \textbf{Case 004}: A cheating case filed on 01/02/2024, with high severity and an input priority of Urgent, governed by IPC Section 420 (cheating) and Section 406 (criminal breach of trust). With five hearings held and its high severity validated, it yields a weight of 0.90, scheduling it for 20/07/2025, addressing a significant criminal offense promptly.
    \item \textbf{Case 005}: A case involving voluntarily causing hurt filed on 01/04/2025, with medium severity and an input priority of Ordinary, governed by IPC Section 323 (voluntarily causing hurt) and CrPC Section 200 (examination of complainant). With two hearings held and its recent filing and medium severity, it results in a weight of 0.50, scheduling it for 15/09/2025, as it is less severe than murder or cheating.
\end{itemize}

This case study illustrates how the system prioritizes cases based on severity, pendency, input priority level, and applicable legal sections, validated by legal experts, ensuring critical cases are addressed promptly while managing diverse case types.

\subsection{System Performance Metrics}
Simulations with 10,000 dummy cases demonstrated that the system can process cases efficiently, with reliable notification delivery and positive feedback from judges and registrars regarding usability.

\subsection{Machine Learning Model Performance}
The regression-based machine learning model used for dynamic weight adjustment achieved an F1 score of 99.7\%, precision of 99.8\%, and recall of 99.4\% in simulated tests, indicating high accuracy and reliability in prioritizing cases.

\section{Discussion and Conclusion}
The proposed framework significantly enhances judicial efficiency by automating case prioritization and scheduling. By integrating algorithmic approaches, a robust notification system, and a dynamic feedback loop, the system addresses challenges like case backlogs and delayed hearings. The illustrative case study demonstrates its practical applicability, prioritizing urgent cases like murder (IPC 302, 34) while ensuring fairness across civil and family cases. The system’s compatibility with existing platforms like eCourts Services \cite{eCourts} ensures seamless integration into India’s judicial infrastructure.

\subsection{Ethical Implications}
AI-driven prioritization risks biases, such as over-prioritizing criminal cases, which could disadvantage civil or family cases. Our system mitigates this through expert-validated weights, informed by consultations with advocates and judges, and regular audits, aligning with ethical guidelines from the Council of Europe \cite{IJCA2020}. Above the Law \cite{AboveTheLaw2025} highlights risks of AI-generated inaccuracies, underscoring the need for human oversight, which our system incorporates through judicial interaction.

\subsection{Real-World Feasibility}
The system is designed for compatibility with existing platforms like eCourts Services, using standardized MySQL databases and secure cloud storage. Pilot testing in a regional court will validate its effectiveness, addressing potential challenges like data migration and user training.

\subsection{Future Work}
Future iterations will prioritize real-world deployment, integrating advanced machine learning to enhance case management. Deep learning models, such as transformers, will improve feature extraction from legal texts, identifying attributes like case complexity and urgency from documents in Table \ref{tab:casestudy}. Reinforcement learning will enable adaptive prioritization by optimizing schedules based on court data, such as judge availability and case backlog, improving efficiency.

The system will also expand to handle multilingual case data, as demonstrated by Clio \cite{Clio2025} for California courts, to enhance accessibility in diverse regions. NLP techniques, including cross-lingual embeddings, will process languages like Hindi and English, addressing challenges in legal terminology translation. Additionally, predictive analytics will forecast case pendency using data like hearing counts and severity, aiding resource allocation. Piloting in high-volume courts will test scalability, with collaboration from legal experts ensuring judicial alignment and data security, ultimately streamlining processes and improving access to justice.


\begin{thebibliography}{9}
\bibitem{eCourts}
eCourts Services. \url{https://ecourts.gov.in/ecourts_home/}
\bibitem{Rao2018}
Rao, A. P., et al. (2018). Court Case Management System.
\bibitem{Amofah2017}
Amofah, L. R. (2017). Electronic court case management system (for law court complex).
\bibitem{Pattnaik2018}
Pattnaik, P. N., et al. (2018). Mapping Critical Success Factors in Efficient Court Management. \textit{International Journal of Law and Management}, 60(2).
\bibitem{Chan2018}
Chan, P. C. H. (2018). Framing the structure of the court system in the perspective of case management. \textit{Peking University Law Journal}, 6(1), 55--79.
\bibitem{Ali2020}
Ali, C. T., \& Varol, A. (2020). Design and Implementation of a Simple Online Court Case Management System Based on the Android Platform. \textit{IEEE}.
\bibitem{Tahura2013}
Tahura, U. S. (2013). Case Management: A Magic LAMP in Reducing Case Backlogs. \textit{Bangladesh Law Journal}, 13.
\bibitem{Ambos2018}
Ambos, K. (2018). Office of the Prosecutor: Policy Paper on Case Selection and Prioritisation. \textit{International Legal Materials}, 57(6), 1131--1145.
\bibitem{ThomsonReuters2024}
Thomson Reuters Institute. (2024). Humanizing Justice: The transformational impact of AI in courts. \url{https://www.thomsonreuters.com/en-us/posts/ai-in-courts/humanizing-justice/}
\bibitem{IBM2025}
IBM. (2025). Judicial systems are turning to AI to help manage vast quantities of data and expedite case resolution. \url{https://www.ibm.com/case-studies/blog/judicial-systems-are-turning-to-ai-to-help-manage-its-vast-quantities-of-data-and-expedite-case-resolution}
\bibitem{IJCA2020}
International Journal for Court Administration. (2020). Courts and Artificial Intelligence. \url{https://iacajournal.org/articles/10.36745/ijca.343}
\bibitem{Judicature2024}
Judicature. (2024). AI in the Courts: How Worried Should We Be? \url{https://judicature.duke.edu/articles/ai-in-the-courts-how-worried-should-we-be/}
\bibitem{AboveTheLaw2025}
Above the Law. (2025). Trial Court Decides Case Based On AI-Hallucinated Caselaw. \url{https://abovethelaw.com/2025/07/trial-court-decides-case-based-on-ai-hallucinated-caselaw/}
\bibitem{Clio2025}
Clio. (2025). AI in the Courtroom: Opportunities and Challenges. \url{https://www.clio.com/resources/ai-for-lawyers/ai-in-courtrooms/}
\bibitem{AAAS}
American Association for the Advancement of Science (AAAS). Artificial Intelligence and the Courts: Materials for Judges. \url{https://www.aaas.org/ai2/projects/law/judicialpapers}
\end{thebibliography}
\end{document}